\documentclass[prl,twocolumn,showpacs,floats]{revtex4}

\usepackage{graphicx}
\usepackage{amssymb}

\begin{document}

\title{Unusual Electronic Structure and Observation of Dispersion Kink in CeFeAsO Parent Compound of FeAs-Based Superconductors }

\author{Haiyun Liu$^{1}$, G. F. Chen$^{2}$, Wentao Zhang$^{1}$, Lin Zhao$^{1}$, Guodong  Liu$^{1}$, T.-L. Xia$^{2}$,
Xiaowen Jia$^{1}$, Daixiang Mu$^{1}$, Shanyu Liu$^{1}$, Shaolong
He$^{1}$, Yingying Peng$^{1}$, Junfeng He$^{1}$, Zhaoyu Chen$^{1}$,
Xiaoli Dong$^{1}$, Jun Zhang$^{1}$,  Guiling Wang$^{3}$, Yong
Zhu$^{3}$, Zuyan Xu$^{3}$, Chuangtian Chen$^{3}$}

\author{X. J. Zhou$^{1,}$}
\email[Corresponding author: ]{XJZhou@aphy.iphy.ac.cn}

\affiliation{
\\$^{1}$National Lab for Superconductivity, Beijing National Laboratory for Condensed Matter Physics, Institute of Physics,
Chinese Academy of Sciences, Beijing 100190, China
\\$^{2}$Department of Physics, Renmin University of China, Beijing 100872, China
\\$^{3}$Technical Institute of Physics and Chemistry, Chinese Academy of Sciences, Beijing 100190, China
}
\date{December 15, 2009}
%
%

\begin{abstract}

We report the first comprehensive high-resolution angle-resolved
photoemission measurements on CeFeAsO, a parent compound of
FeAs$-$based high temperature superconductors with a
mangetic$\slash$structural transition at $\sim$150 K. In the
magnetic ordering state, four hole-like Fermi surface sheets are
observed near $\Gamma$(0,0) and the Fermi surface near
M($\pm$$\pi$,$\pm$$\pi$) shows a tiny electron-like pocket at M
surrounded by four strong spots. The unusual Fermi
surface topology deviates strongly from the band structure
calculations.  The electronic signature of the
magnetic$\slash$structural transition shows up in the dramatic
change of the quasiparticle scattering rate.  A dispersion kink at
$\sim$25 meV is for the first time observed in the parent compound
of Fe-based superconductors.

\end{abstract}


\maketitle

The recent discovery of Fe-based
superconductors\cite{Kamihara,ZARenSm,RotterSC,MKWu11,CQJin111} has
attracted much attention because it represents the second class of
high-T$_c$ superconductors after the first discovery of
high-T$_c$ copper-oxide superconductors. The parent
compound is antiferromagnetic\cite{PCDai,BFSNeutron3}; superconductivity is
realized by suppressing the magnetic ordering through chemical
doping\cite{Kamihara,ZARenSm,RotterSC,MKWu11,CQJin111} or applying
high pressure\cite{PressureTc}. Understanding the nature and origin
of the magnetic ordering is thus crucial in
understanding the interplay between magnetism and superconductivity
in the Fe-based superconductors.  Among various Fe-based families\cite{Kamihara,ZARenSm,RotterSC,MKWu11,CQJin111}, the
LnFeAsO system (so-called 1111-type with Ln representing rare
earth elements) is particularly interesting because of its highest
T$_c$ ($\sim$55 K)\cite{ZARenSm} which remains the only family with
T$_c$ above 40 K so far.  Despite being the first Fe-based
superconductors discovered\cite{Kamihara}, the 1111-type compounds
have not been studied extensively due to the difficulty in obtaining
high quality and sizable single crystal samples. While a number of
angle-resolved photoemission (ARPES) measurements have been
performed on the (Ba,Sr)Fe$_2$As$_2$ parent compounds (so-called
122-type)\cite{LXYangBFA,CLiuBFeAs,HYLiuSrKFA,VBZ_BFA,ZHasanSFS,GDLiu122,MYiLowT},
very few detailed ARPES studies have been carried out on the
FeAs-based 1111-type compounds\cite{Kaminski1111,DHLuPhysicaC}.
Investigation of the 1111 family is important not only to understand
its unique high T$_c$ but also to sort out universal features among
various families that are generic for high-T$_c$ superconductivity
in the Fe-based superconductors.

\begin{figure}[tbp]
\includegraphics[width=0.96\columnwidth,angle=0]{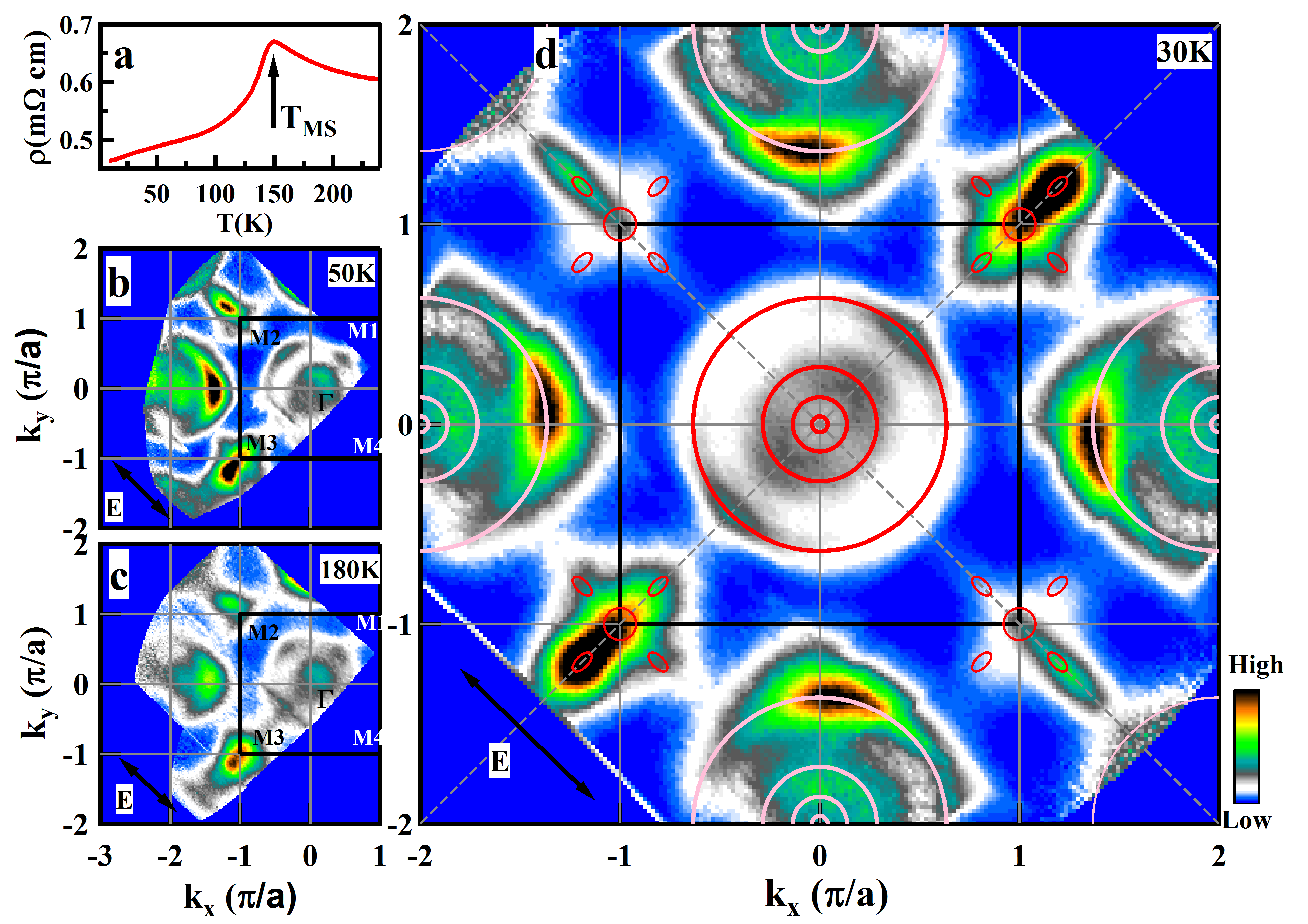}
\caption{(a). Resistivity$-$temperature dependence of CeFeAsO single crystal.
(b). Spectral intensity as a function of k$_x$ and k$_y$ integrated over a [-5 meV,5 meV]
energy window with respect to the Fermi level at 50 K. (c). Fermi surface mapping at
180 K on the same sample as (b). The incident light polarization is marked as the
black arrows. (d). Fermi surface mapping at 30 K from another CeFeAsO sample.}
\end{figure}

In this paper we report first comprehensive high resolution ARPES measurement on CeFeAsO.
Unusual Fermi surface topology in the magnetic ordering state is observed: (1). Four
hole-like Fermi surface sheets are observed near $\Gamma$(0,0) that
are different from three sheets expected from band structure
calculations; (2). The Fermi surface near M($\pm\pi$,$\pm\pi$) is
composed of a tiny electron-like Fermi pocket at M  surrounded by
four strong spots that is dramatically different from large electron-like Fermi surfaces expected from band
structure calculations.  The electronic signature of
the magnetic$\slash$structural transition  shows up in CeFeAsO in
the obvious change of the quasiparticle scattering rate. For the
first time we have observed the development and enhancement of
dispersion kink at $\sim$25 meV in the magnetically-ordered state.
These results will shed key light on understanding the magnetism
and superconductivity in the Fe-based superconductors.

The ARPES measurements were carried out using our lab system equipped
with Scienta R4000 analyzer and VUV5000 UV source which gives a
photon energy of Helium I at h$\upsilon$= 21.218
eV\cite{GDLiuARPES}. The overall energy resolution is 10 meV
and the angular resolution is $\sim$0.3 degree.  The CeFeAsO
single crystals with a size of 1$\sim$2 mm were grown using flux method. The
resistivity-temperature dependence is nearly identical to that of
the polycrystalline CeFeAsO sample\cite{GFChenCe} with a
magnetic$\slash$structural transition temperature (T$_{MS}$) at
$\sim$ 150 K (Fig. 1a).  The crystals were cleaved {\it in situ} and measured
in vacuum with a base pressure better than 5$\times$10$^{-11}$ Torr.

\begin{figure}[tbp]
\includegraphics[width=0.70\columnwidth,angle=0]{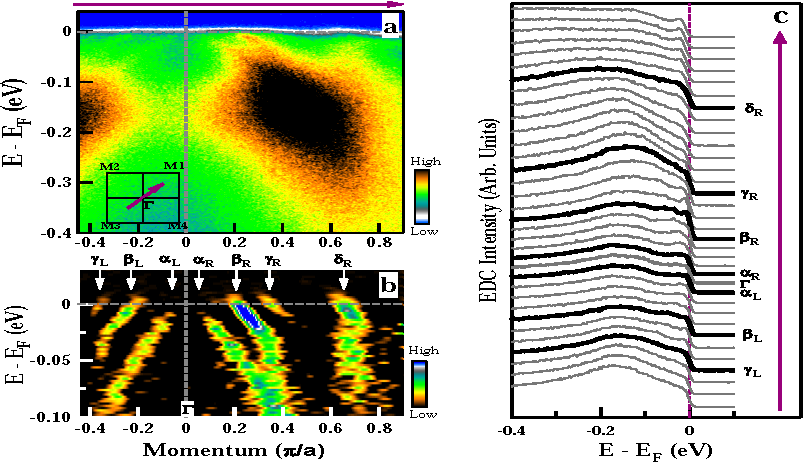}
\caption{Electronic structure of CeFeAsO near $\Gamma$ measured at
20 K. (a). Original photoemission image measured on a $\Gamma$-M cut
as shown in the inset. (b). Second derivative image obtained from
(a) with all Fermi crossings marked.  (c). Corresponding EDCs with those at k$_F$ marked as thick lines.
}
\end{figure}


Figure 1 shows the Fermi surface mapping of CeFeAsO covering a
multiple Brillouin Zones (BZs). Figs. 1b and c are original Fermi
surface data measured below and above T$_{MS}$$\sim$150 K,
respectively, while Fig. 1d is obtained by symmetrizing the original
data under four-fold symmetry measured on another sample. The
spectral intensity in the 2nd BZ is stronger than
that of the first one. Also the overall spectral distribution lacks
strict four-fold symmetry around $\Gamma$(0,0) which may be caused
by the photoemission matrix-element effect. Detailed analysis of the
electronic structure near $\Gamma$, M3(-$\pi$,-$\pi$), and
M2(-$\pi$,$\pi$) are shown in Fig. 2, Fig. 3 and Fig. 4,
respectively (we denote the four M
points as M1, M2, M3 and M4, as shown in Figs. 1b and c).

\begin{figure*}[floatfix]
\includegraphics[width=1.8\columnwidth,angle=0]{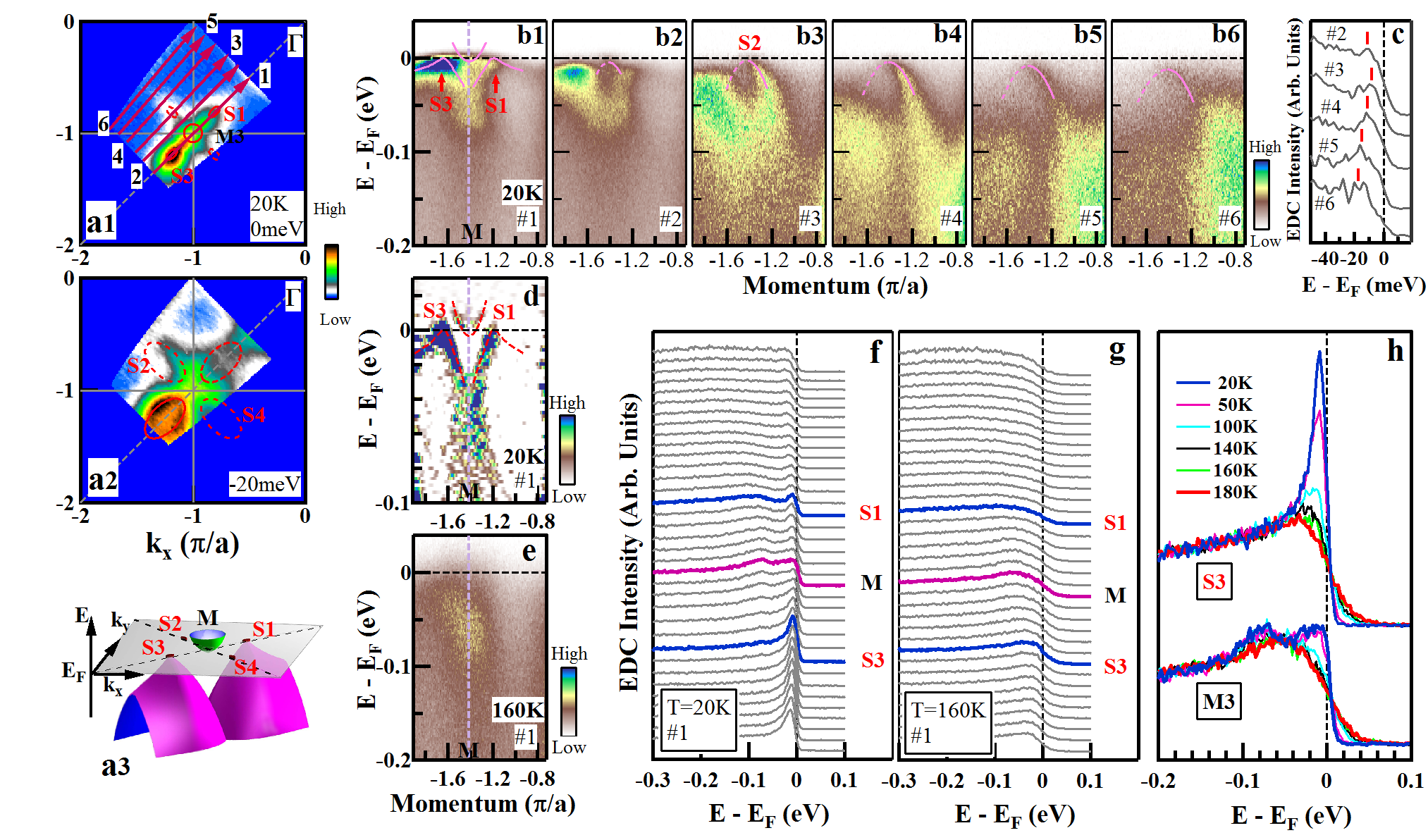}
\caption{Electronic structure of CeFeAsO near M3(-$\pi$,-$\pi$).
(a). Spectral intensity as a function of k$_x$ and k$_y$
integrated over [-1 meV,1 meV](a1) and [-21 meV,-19 meV](a2) with respect to the
Fermi level measured at 20 K. The strong spot S3 in (a1)
increases in area with increasing binding energy to become hole-like
Fermi pocket in (a2). (a3). Schematic drawing of band structure
around M. For clarity, we show only the $\Gamma$-M3-parallel line
patch; the $\Gamma$-M3-perpendicular patch is similar but not drawn here.
(b). Photoemission images taken along 6 typical momentum cuts with
their location shown in (a1). Pink lines are
guides to the eye for the band structure. (c). EDCs near the band
top for cuts $\#2$ (Fig. b2) to $\#$6 (Fig. b6). The EDC for the cut
$\#$3 is closest to the Fermi level while the others are below the
Fermi level. (d). Second derivative image from Fig. 3b1 for the cut
$\#$1. Red dashed curves denote two hole bands (S1 and S3) and an
electron band at M. (e). Photoemission image of the cut $\#$1 at 160
K. (f) and (g) show EDCs of the cut $\#$1 at 20 K and 160 K
(corresponding to Fig. 3b1 and 3e). (h). Temperature dependence of
EDCs at S3 and M.}
\end{figure*}

The Fermi surface of CeFeAsO consists of four hole-like sheets near
$\Gamma$ in the magnetic-ordering state (Fig. 1d). This can be seen
clearly from the band structure measurement along the $\Gamma$-M cut
(Fig. 2a and 2b) where four hole-like bands are identified (denoted
as $\alpha$, $\beta$, $\gamma$, and $\delta$ in Fig. 2b). The
corresponding photoemission spectra (energy distribution curves,
EDCs, Fig. 2c) indicate that these four bands all cross the Fermi
level with the Fermi momenta (k$_F$) of 0.04 ($\alpha$),
0.22($\beta$), 0.35 ($\gamma$), and 0.66 ($\delta$) in a unit of
$\pi$/a (lattice constant a=3.996 $\AA$).  The Fermi surface volume
calculation gives an electron counting of 2 ($\alpha$), 1.9 ($\beta$), 1.8 ($\gamma$) and 1.39 ($\delta$).
The Fermi velocities of these four bands are 0.34($\alpha$),
0.30($\beta$), 0.23($\gamma$) and 0.46 ($\delta$) in a unit of
eV$\cdot$${\AA}$. In addition to these four bands, there is also a
flat band at $\sim$-0.18 eV as seen from the EDCs in Fig. 2c.

The observation of four hole-like Fermi surface sheets near $\Gamma$
is surprising because only three bands are expected from
band structure calculations either in non-magnetic
state\cite{DJSingh1111} or magnetic state\cite{FJMa}. This is
different from the previous measurement on LaFeAsO where three bands
are resolved near $\Gamma$ with only two of them crossing the Fermi
level\cite{DHLuPhysicaC}. It is also different from the 122-type
parent compound BaFe$_2$As$_2$ where only three bands near $\Gamma$
are observed below T$_{MS}$\cite{GDLiu122}.  However, we note that
the inner three bands of CeFeAsO ($\alpha$, $\beta$ and $\gamma$)
show strong resemblance to those in BaFe$_2$As$_2$ \cite{GDLiu122}
with even similar Fermi momenta and the outer large Fermi surface
($\delta$) is similar to that observed in LaFeAsO\cite{DHLuPhysicaC}
and LaFePO\cite{DHLuLaFePO}. One immediate possibility is
whether the extra band(s) in CeFeAsO is due to band folding
associated with the magnetic$\slash$structural transition. We believe
this is unlikely because the folding effect in CeFeAsO is weak since
there is no similar hole-like bands observed near M, as we will see
later in Fig. 3 and Fig. 4. In particular, the large $\delta$ band
can not be due to such a folding effect because it is present both
below and above T$_{MS}$ (Figs. 1b and c).   The other possibility is
whether the extra band is specific for CeFeAsO because of its Ce 4f
electrons.  We performed band structure calculations 
and found that for CeFeAsO there appear two extra large hole-like Fermi
surface sheets near $\Gamma$ from the Ce-related electronic states.
The fact that we see only one ($\delta$) and its strong similarity to that in LaFeAsO\cite{DHLuPhysicaC} and
even LaFePO\cite{DHLuLaFePO} makes us believe it is more likely a
generic feature for the 1111-type compounds. The third possibility is whether the extra
band(s) could be due to surface state. Considering the inner three
bands are similar in CeFeAsO and BaFe$_2$As$_2$ systems, it comes to
whether the large Fermi surface $\delta$ could be surface state
common in 1111-type compounds.  We note that a simple electron
counting of the inner three Fermi surface ($\alpha$, $\beta$ and
$\gamma$) yields a total of 5.7 that is close to the expected value
of 6, while the electron counting of all four sheets is 7 which is
apparently larger than 6. This seems to suggest not all of four
Fermi surface sheets are intrinsic that are representative of the
bulk property.



The Fermi surface mapping of CeFeAsO near M (Fig. 1) is
characterized as two line patches crossing each other at M: one
patch is along the $\Gamma$-M direction with strong spectral weight
in the second BZ, while the other weak patch
perpendicular to the $\Gamma$-M direction. The two independent
measurements near M2 and M3 give similar results (Fig. 1d). Similar
cross-shaped pattern can also be seen near M in
LaFeAsO\cite{DHLuPhysicaC}. To understand the formation of these
line patches, we carried out detailed momentum-dependent
measurements for the $\Gamma$-M-parallel patch (Fig. 4b1-6 and Fig.
4c1-6) as well as for the $\Gamma$-M-perpendicular patch (Fig.
3b1-6). It turns out that, in both cases, the band structure along
the line patch shows a series of hole-like inverse-parabolic
bands; the line patch in the Fermi surface mapping
then corresponds to the spectral weight from the top of these bands.
The energy position of the band top shows a maximum on each side of
M that is closest to the Fermi level, as seen from Fig. 3b, Fig. 4b
and 4c, and more clearly from the EDCs in Fig. 3c. These give rise
to four strong spots (S1, S2, S3 and S4 in Figs. 1c, 3a and 4a)
around M.  In addition, a careful inspection of bands right near the
M point suggests the existence of a tiny electron-like pocket at M.
This can be seen from the momentum cut along the $\Gamma$-M3 line
(Fig. 3b1 and Fig. 3d for the cut $\#1$ in Fig. 3a). Fig. 3a3 shows
a schematic drawing of the band structure near M along one line patch.
The peculiar Fermi surface topology near M deviates significantly
from the band structure calculations which predict two crossed electron-like
ellipses\cite{DJSingh1111,FJMa}. It is also distinct from that
observed in BaFe$_2$As$_2$\cite{GDLiu122} but bears some
resemblance to that of optimally-doped
(Ba$_{0.6}$K$_{0.4}$)Fe$_2$As$_2$\cite{LZhao122,VBZ_BFA,Hasan122}.

The electronic signature of magnetic$\slash$structural transition in
CeFeAsO shows up in the quasiparticle scattering rate. This can be
seen from the temperature evolution of the photoemission spectra at
M (EDCs at M3 in Fig. 3h) and at the strong spot S3 (Fig. 3h and
Fig. 4f).  The EDCs at M3 (Fig. 3h) show a broad peak above T$_{MS}$
and evolve into two-peak structure at low temperature with
sharpening of the peak near the Fermi level.  The change of EDCs
near the  strong spot S3 (Fig. 3h and Fig. 4f) is more dramatic: it
evolves from a broad feature above T$_{MS}$ into a sharp
quasiparticle peak at low temperature in the magnetic ordering
state. Our simulations indicate that such dramatic sharpening cannot
be attributed to trivial temperature broadening in terms of Fermi
distribution. Therefore, this dramatic decrease of the quasiparticle
scattering rate is a clear signature of the magnetic transition.

\begin{figure*}[tbp]
\includegraphics[width=1.8\columnwidth,angle=0]{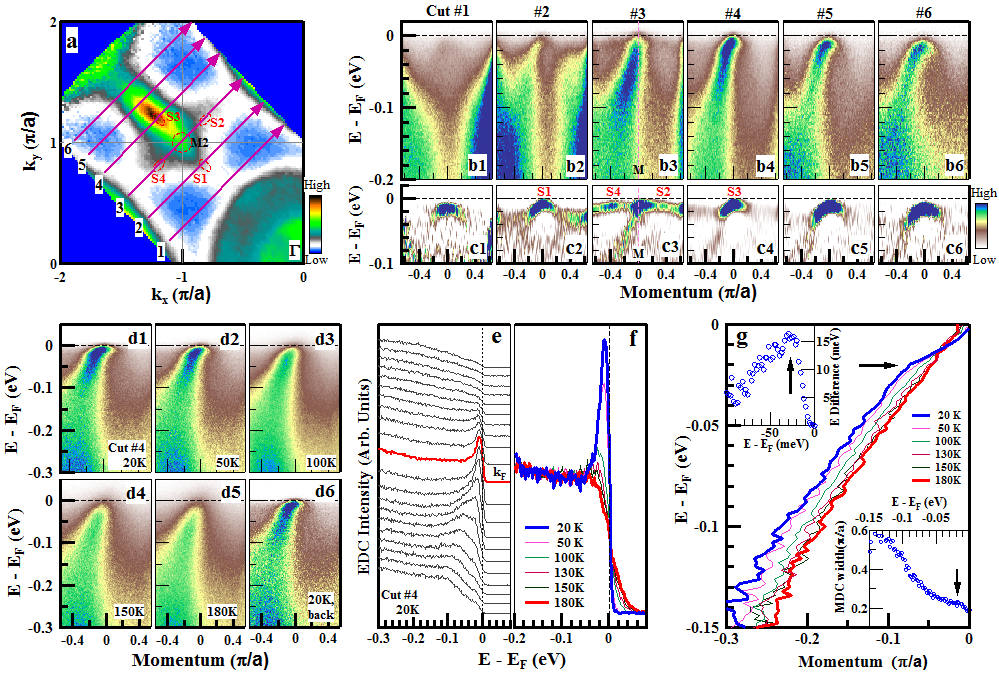}%
\caption{Electronic structure of CeFeAsO near M2(-$\pi$,$\pi$). (a).
Spectral intensity as a function of k$_x$ and k$_y$ integrated over
[-10 meV, 10 meV] energy window with respect to the Fermi level
measured at 30 K. (b). Photoemission images at 6 typical momentum
cuts with their location marked in Fig. 4a. (c). Corresponding
second derivative images of Fig. 4b. (d). Photoemission images of
the cut $\#4$ measured at different temperatures, warming up from 20
K (d1) to 180 K (d5) and then cooling back to 20 K (d6). (e). EDCs
of the cut $\#4$ at 20 K corresponding to Fig. 4b4. (f). Temperature
dependence of the EDCs at k$_{F}$ of the cut $\#4$. (g).
Quantitative dispersions of the cut $\#4$ at different temperatures
extracted by fitting MDCs from photoemission images in Fig. 4d. A
kink at $\sim$ 25 meV at low temperature is marked by the black
arrow. The upper left inset shows the energy difference between the
measured dispersion at 20 K and a straight line connecting two points
on dispersion at E$_F$ and -0.15 eV. The corresponding MDC width at
20 K is shown in the lower right inset.}
\end{figure*}


As seen in Fig. 4,  the energy bands near the strong spots (Fig.
4b4-5 near the strong spot S3) show a clear kink in the dispersion.
Quantitative MDC (momentum distribution curve) analysis (Fig. 4g)
indicates that the kink in dispersions occur at $\sim$25 meV,
accompanied by a drop in the MDC width (bottom-right inset of Fig.
4g). The observations clearly indicate the existence of an energy scale at 25 meV which
are reminiscent of the electron coupling with bosonic
modes as seen in cuprate superconductors\cite{CuprateCoupling}. One
possibility is that the kink is due to electron coupling to
phonons.  This scenario is consistent with the calculation of the
Eliashberg function $\alpha$$^{2}$F($\omega$) associated with
electron-phonon coupling in LaFeAsO that exhibits an obvious peak
near 25 meV\cite{LBoeri}. Raman scattering also shows the existence
of a particular B$_{1g}$ phonon mode at $\sim$25 meV associated with
the Fe vibrations\cite{VGHRaman}. We remark that this does not mean
there are no other modes existing between  0 and 25 meV because the
manifestation of multiple modes coupling may give rise to a similar
kink at 25 meV\cite{XJZhou}. Although electron-phonon coupling provides
a likely origin of the kink, we cannot fully rule out other
possibilities of electronic or magnetic origins. A hint of a
$\sim$25 meV kink can also be seen near $\Gamma$ in the $\beta$ band
(Fig. 2b), but not clear in the $\alpha$ band, suggesting its
orbital-selective nature.   As seen in the temperature
dependence measurements (Fig. 4d and 4g), the kink develops
below T$_{MS}$ and gets enhanced at low temperatures, which indicates
that the kink is closed related to the magnetic$\slash$structural
transition.  This is distinct from the kink reported in the
superconducting (Ba$_{0.6}$K$_{0.4}$)Fe$_2$As$_2$ near the $\Gamma$
point which is considered to be related to superconductivity\cite{Hasan122,HDingKink}.


In summary, from our comprehensive ARPES measurements we find that
CeFeAsO exhibits unusual electronic structure that deviates strongly
from the band structure calculations.  The electronic signature of
the magnetic$\slash$structural transition shows up in the dramatic
change of the quasiparticle scattering rate. We have identified for
the first time a dispersion kink in this parent compound which is
closely related to the magnetic$\slash$structural transition.  These
results provide a clear indication on the interplay between charge,
spin and lattice in the FeAs-based parent compounds. These rich
information also provide an opportunity to compare and contrast with
other compounds in order to sort out universal features responsible
for the FeAs-based superconductors. To better understand the unusual
behaviors of these materials, further work needs to be done to well
characterize the sample surface, particularly the surface doping or
surface structure reconstruction.


We thank D.-H. Lee, T. Xiang, G.-M. Zhang, H. Zhai and Z. Y. Lu for
helpful discussions. This work is supported by the NSFC, the MOST of
China (973 project No: 2006CB601002, 2006CB921302, 2009CB929101) and
CAS.

\end{document}